\documentclass[twocolumn,final,aps,prl,superscriptaddress]{revtex4}
\usepackage{amsmath}
\usepackage{pst-node,graphicx,array}
\usepackage{pstricks}

\begin{document}

\title{Charge Ordering in the $Pt_2(dta)_4I$ Compound}

\author{Marie-Bernadette LEPETIT}
\email{marie@irsamc.ups-tlse.fr}
\author{Vincent ROBERT}  
\affiliation{Laboratoire de Physique Quantique, IRSAMC/UMR 5626, Universit\'e Paul Sabatier, 118 route de Narbonne, F-31062 Toulouse, Cedex 4, FRANCE} 

\date{\today}

\begin{abstract}
The present paper tries to elucidate the electronic structure of the
 $Pt_2(L)_4I$ family of compounds in the semi-conducting and metallic
 phases, using correlated {\em ab initio} embedded fragment
 calculations. The usual model associating the different phases with
 different intra-dimer charge orders, is energetically ruled out by
 our results. A careful analysis of the local low-energy degrees of
 freedom lead us to propose a mechanism associating charge transfer
 from the iodine $5p_z$ orbital with the vibrational degrees of
 freedom of the $dta$ ligands. It is shown that such a model can
 account for both the XPS and IR experimental results at the phase
 transition.

\end{abstract}

\maketitle


Low-dimensional compounds have attracted a lot of interest in both
experimental and theoretical communities for the last twenty years.
Among these, the one-dimensional (1D) halogen-bridged transition-metal
complexes exhibit several very peculiar properties which have induced
a lot of research work. One can cite, for instance, large third-order
non-linear optical properties~\cite{mx:isawa}, photo-generation and
annihilation processes of mid-gap states~\cite{mx:wada}, or intense
dichroic inter-valence charge transfer absorption~\cite{mx:okamoto}.
These systems are perfectly one-dimensional, mixed valence and
strongly correlated, being thus paradigms of strongly interacting 1D
systems, and exhibiting a large number of different phases.  The
nature of the ground state can be controlled not only by pressure and
temperature but also by chemical substitution of either transition
metal atoms, halogen atoms or even counter
ions~\cite{mx:cions}. However all the so-called $MX$ ($M$ metal, $X$
halogen) systems which have been synthesized so far are insulators or
semi-conductors. Extensive works have been done without success, on
these systems, in order to induce metalization under
pressure~\cite{mx:metal}. Therefore researches have
switched focus toward the synthesis of bi-nuclear~\cite{mmx:synth}
(and very recently, tetra-nuclear~\cite{m4x:synth}) $MMX$ systems. The
$P\!t_2(dta)_4I$ complex ($dta$=dithioacetate=$CH_3CS_2^-$) has thus
become the second example (following $KCP(B\!r)$~\cite{KCP(Br)}) of 1D
system exhibiting metallic transport without involving a $\pi$
electronic system.  This system has since then become the archetype of
a series of similar $MMX$ intriguing compounds.

\begin{figure}[h]
\resizebox{3.5cm}{3.5cm}{\includegraphics{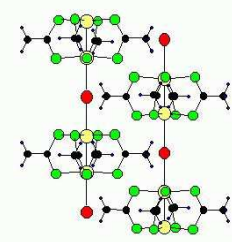}}
\caption{Crystal structure of the $P\!t_2(dta)_4I$ compound at
$298K$~\cite{mmx:synth}. One should note that the two $P\!tS_4$ units
of the $P\!t_2 (dta)_4$ dimer are twisted from the eclipsed position
with an average angle of $21^\circ$.}
\label{fig:struct}
\end{figure}

The $P\!t_2(dta)_4I$ compound has a neutral chain structure consisting
of dimeric units $\left[P\!t_2 (dta)_4\right]$ bridged through halides
(see figure~\ref{fig:struct}). 
From electrical conductivity and thermoelectric power
measurements~\cite{mmx:jacs} three different phases have been
identified in the range of temperatures from $90K$ to $450K$.  The
system is a semi-conductor between $90K$ and $T_{MS}=300K$. From
$300K$ to $373K$, the system is metallic and then undergoes a metal to
metal phase transition at $T_{MM}=373.4K$. 

Let us now analyze the electronic structure. A formal charge
analysis leads to the $(P\!t)_2^{5+}(dta)_4^{4-}I^-$ charge transfer
form. Ligand field analysis shows that the Fermi level orbitals are
the $P\!t$ non-degenerated $5d_{z^2}$ orbitals, $z$ being the axis of
the chains. It results that the low-energy physics of the
$P\!t_2(dta)_4I$ compound can be described as a correlated
three-quarter filled band supported by the platinum $5d_{z^2}$
orbitals. Such a mixed-valence system can be subject to a large number
of valence instabilities associated with different charge orderings,
namely one can find
\begin{enumerate}  \itemsep -0.8ex
\item[(A)] $\rm
I^-\!....\;P\!t^{2.5+}\!..\;P\!t^{2.5+}\!....\;I^-\!....\;P\!t^{2.5+}\!..\;P\!t^{2.5+}\!....\;I^-$
\item[(B)] $\rm I^-\!....\;P\!t^{2+}\!..\;P\!t^{3+}\!...\;I^-\!....\;P\!t^{2+}\!..\;P\!t^{3+}\!...\;I^-$ 
\item[(C)] $\rm I^-\!....\;P\!t^{2+}\!..\;P\!t^{3+}\!...\;I^-\!...\;P\!t^{3+}\!..\;P\!t^{2+}\!....\;I^-$ 
\item[(D)] $\rm I^-\!...\;P\!t^{3+}\!..\;P\!t^{3+}\!...\;I^-\!....\;P\!t^{2+}\!..\;P\!t^{2+}\!....\;I^-$ 
\end{enumerate}
Based on several experimental results, different authors have tried to
interpret the electronic structure of the different observed phases
according to the above charge orderings.

In the $150-300K$ temperature range, XPS experiments~\cite{mmx:jacs}
exhibit two signals associated respectively with $P\!t^{2+}$ and
$P\!t^{3+}$. Similarly, IR~\cite{mmx:jacs} spectra show a doubling of
the $C=S$ stretching mode upon cooling from the metal (M) to the
semi-conductor (SC) phase. This double peak has been interpreted as a
signature of a double valency of the platinum atoms in the
SC phase, therefore excluding the electronic structure
pictured in (A).  As for magnetic susceptibility measurements, a
slight increase is observed upon cooling, below $T_{MS}$.  Kitagawa
{\it et al}~\cite{mmx:jacs} therefore concluded to a zero spin gap and
a $4k_F$-CDW Mott-Hubbard ground state for the SC phase, that is an
electronic structure of type (B).  However, it should be pointed out
that i) the XPS experiments have a very short time scale ($\sim
10^{-17}s$) and therefore cannot differentiate between mixed and
averaged valencies (as stated in references~\cite{mmx:jacs}
and~\cite{xps:calo}), ii) the XPS $P\!t^{2+}$ and $P\!t^{3+}$ signals
intensities exhibit a 3 to 1 ratio, incompatible with any of the
above-cited charge orderings.

The aim of this paper is to analyze the above issue,
using {\em ab initio} quantum chemical calculations.  We used for this
purpose the embedded fragment technique associated with {\em ab
initio} spectroscopy methods, that is large atomic basis
sets~\cite{mmx:base} and configurations interaction techniques
dedicated to the accurate evaluation of both transitions energies and
states wave-functions.  More precisely we used the Difference
Dedicated Configuration Interaction (DDCI) method~\cite{ddci,casdi}. It
this method the orbital space is split into three sets~: {\bf the
occupied orbitals} that are always doubly-occupied in the system, {\bf
the active or magnetic orbitals} that have fluctuating occupation
numbers or spins in the ground and excited states, {\bf the virtual
orbitals} that are always empty in the system. The DDCI method is an 
exact diagonalization method on a selected configuration space
designed as follow :
\begin{itemize}
\item all configurations that can be built from the above occupation
rules, this is the Complete Active Space (CAS) which is used as a
reference space and corresponds to the support of the model
Hamiltonian,
\item then on this CAS, all single excitations and all double
excitations participating in the excitation energy at the second order
of perturbation (that is the configurations responsible for the
screening effects on the interactions within the CAS).
\end{itemize}
This approach has been very successful in the quantitative evaluation
of quantities such as magnetic couplings, charge order, spin
polarization, etc.  both in molecular magnetic systems and in strongly
correlated materials (see for instance
references~\cite{molmagn,Xino,vana3}). Its accuracy can be exemplified
on the high-$T_c$ superconductor $N\!d_2C\!uO_4$ for which the
effective exchange coupling has been computed within experimental
accuracy ($J_{calc} = -126.4$meV~\cite{Xino},
$J_{exp}=-126\pm5$meV~\cite{cu:j}).  The success of these embedded
fragment methods in transition metal compounds is due i) to the
locality of the $d$ Fermi electrons responsible for the low-energy
properties~\cite{revue} and ii) to the possibility of using quantum
chemical methods able to accurately treat open-shell problems and
screening effects.

The fragment used in the present calculations is composed of one
$P\!t_2(dta)_4$ dimeric unit and its neighboring iodine atoms, that is
$IP\!t_2(dta)_4I$. Its geometry has been extracted from the X-ray
structure at room temperature~\cite{mmx:synth}. This fragment was
embedded in a large bath of total-ion
pseudo-potentials~\cite{mmx:base} (TIP) and point charges, designed in
order to account for the Madelung potential and the short-range
exclusion and exchange effects of the rest of the crystal. We have
considered the three different possible fillings for the platinum
dimer, i.e.  $(P\!t_2)^{6+}$, $(P\!t_2)^{5+}$ and $(P\!t_2)^{4+}$
involved in the charge orderings (A) to (D) ---~that is
$\left(IP\!t_2(dta)_4I\right)^{0,1-, 2-}$.

Results on the $(P\!t_2)^{n+}$ systems ($n=4,5,6$) exhibit a very
low-lying bonding $\sigma_d =\left( d_{z^2}(P\!t_1) +
d_{z^2}(P\!t_2)\right) / N_{\sigma}$ orbital (where $1$ and $2$ refer to
the two $P\!t$ atoms of the dimer and $N_{\sigma}$ is the normalization
constant) and a Fermi level anti-bonding ${\sigma_{d}}^*= \left(
d_{z^2}(P\!t_1) - d_{z^2}(P\!t_2)\right)/N_{\sigma^*}$ orbital. These two
orbitals are separated by the $(dta)$-ligand bonding and non-bonding
orbitals, as well as all the other $P\!t(5d)$ orbitals combinations. 

{ $\bf (P\!t_2)^{5+}$}. 
The intra-dimer charge localization within the $(P\!t_2)^
{5+}$ dimer, as pictured in modes (B) and (C), necessitates the
excitation of an electron from the $\sigma_{d}$ orbital toward the
${\sigma_{d}}^*$ one. Indeed, $\sqrt{2}\; |d_{z^2}\!(P\!t_2)
d_{z^2}\!(P\!t_1) \overline{d_{z^2}\!(P\!t_1)}\rangle = |\sigma_{d}
{\sigma_d}^{*} \overline{\sigma_{d}}\rangle + |\sigma_{d}
{\sigma_{d}}^* \overline{{\sigma_{d}}^*}\rangle$.  However, the
$\sigma_{d} \; \longrightarrow \; {\sigma_{d}}^*$ excitation
requires quite large an energy ($\simeq 5 eV$) and is therefore
strongly hindered.  Coherently, we found the electronic density to be
highly delocalized between the two platinum atoms on the $(P\!t_2)^
{5+}$ system, despite the twisting of the $(dta)$ ligands which is
usually considered as responsible for intra-dimer charge localization
$P\!t^{2+} P\!t^{3+}$~\cite{mmx:vR}. Indeed, the differential
M\"ulliken population between the two $P\!t$ atoms is $ch(P\!t_1) -
ch(P\!t_2) = 0.02 \bar e$ (in full agreement with our periodic mean-field
calculations based on the experimental crystal
structure~\cite{mmx:synth} which yield $0.03\bar e$). These results
therefore exclude the possibility of a charge localization within the
platinum dimer. In conclusion, the description of the electronic
structure of the semi-conducting phase as intra-dimer charge ordered
(whether following the (B) or (C) patterns) is incompatible with 
{\em ab initio} calculations.

From the above results we see that the bonding $\sigma_d$ orbitals on
the $(P\!t)_2(dta)_4$ dimers are not pertinent for the low-energy
physics of the compound. The pertinent model should be a half-filled
one-band (extended) Hubbard model (EHM) based on the dimer
${\sigma_{d}}^*$ orbitals.  Such a model would be in agreement with
the thermoelectric power experiments. The physics of the
one-dimensional EHM is well known~\cite{EHM} and unfortunately does not
exhibit any metallic phase for positive one-site repulsion $U$ and
first neighbor repulsion $V$. One should therefore invoke other
degrees of freedom than the one supported by the $d_{z^2}$ orbitals of the
platinum atom in order to understand the different phases of this
compound.

Let us first examine the low-energy electronic degrees of freedom of
the $(P\!t)_2(dta)_4$ dimer. The first excited state of both the
$(P\!t_2)^{5+}$ and $(P\!t_2)^{6+}$ dimers is a ligand to metal single
excitation where a non-bonding $\pi$ electron of the ligand is excited
into the ${\sigma_d}^*$ orbital. The computed excitation energies are
$2.01eV$ and $0.76eV$ for the $5+$ and $6+$ systems respectively. 

The low-energy physics will be governed by the energy scale of the
inter-dimer interactions. Even-though the $(P\!t)_2(dta)_4\, I \,
(P\!t)_2(dta)_4$ system is much too large to be investigated using the
DDCI methods, the energy scale of the inter-dimer interactions can be
evaluated using qualitative methods. Periodic Hartree-Fock
calculations yield a band width of $0.79eV$. The coulombic repulsion
of two electrons on the dimer ${\sigma_d}^*$ orbital $U_{\sigma_d^*}$
can be roughly evaluated by computing
$$ E\left((P\!t_2)^{4+}\right) + E\left( (P\!t_2)^{6+}\right) -
2\,E\left( (P\!t_2)^{5+}\right) $$ which yields $
U_{{\sigma_d}^*}\simeq 3.9eV$. It results that the effective hopping
$t$ between the ${\sigma_d}^*$ orbitals of two dimers should be of the
order of magnitude of $0.79/4 \simeq 0.2eV$, while the effective
exchange should be of the order of $4t^2/U_{\sigma_d}^* \simeq
0.04eV$. These estimations completely scale out the possible
mobilization of the $2.01eV$ first state on $(P\!t_2)^{5+}$. The
$0.76eV$ first state on $(P\!t_2)^{6+}$ is also excluded since it is
of a different approximate symmetry than the local ground state ($\pi$
instead of $\sigma$) and thus cannot couple in a significant way with
the low-energy band. 

The only other possible electronic degree of freedom involves the
transfer of a $5p_z$ electron of the iodine toward the ${\sigma_d}^*$
orbital of the dimeric unit.  The possible importance of such
excitations has already been pointed out by
S. Yamamoto~\cite{mmx:iode}.
In order to accurately check their pertinence, one should be able to
evaluate the iodine $5p_z$ to ${\sigma_d}^*$ excitation energy and to
compare it to the effective transfer integral between these
orbitals. Unfortunately it is impossible to accurately evaluate these
values using {\em ab initio} methods, since one should be able to
compute excitations on the $I\, (P\!t)_2(dta)_4\, I \,
(P\!t)_2(dta)_4\,I$ system, a presently out of reach problem. However,
it is possible to get a rough estimate of the
$\varepsilon_{{\sigma_d}^*} - \varepsilon_I$ orbital energy difference
between the ${\sigma_d}^*$ orbital and the $p_z$ orbital of the
iodine, as well as of the transfer integral between the iodine $5p_z$
orbital and the dimer ${\sigma_d}^*$ orbital using mean-field
Hartree-Fock theory. It comes $\varepsilon_{{\sigma_d}^*} - \varepsilon_{p_z}
\simeq 0.79eV$ and $t_{{\sigma_d}^*,p_z} \simeq 0.56eV$.  From these
parameters it is possible to compute, using a simple two-bands
tight-binding model, the charge transfer from the iodine $5p_z$ orbital
toward the ${\sigma_d}^*$ orbital of the dimer. The computed charge
transfer is quite large with 0.89 electron. The correlation effects,
that are expected to be quite important in these systems, will
probably reduce this value.  However, it is reasonable to still expect
a significant charge transfer between the two orbitals.

One of the important consequences of such a two-band model is that it
becomes easy to explain the XPS experimental results for the
$P\!t_2I(dta)_4$ compound. Indeed, the total charge carried by the
platinum atoms is now allowed to fluctuate. An average electron
transfer of $\eta$, from the iodine to the platinum dimer will change
the average charge of the latter from $2.5$ to $2.5-\eta$, in
consequence of which the XPS experiments will see $0.5+\eta$
$P\!t^{2+}$  and $0.5-\eta$ $P\!t^{3+}$ ions. The experimental ratio
between the $P\!t^{2+}$ and $P\!t^{3+}$ signals, which is close to
$3:1$, may be understood along this picture with $\eta \simeq 1/4$.
Such a value corresponds to an electronic structure mixing in equal
weights, i) the configuration where the iodine is always doubly occupied 
and ii) the set of singly-excited determinants where an electron has been
transferred from an iodine atom toward a platinum dimer.

We have examined all electronic low energy degrees of freedom, let us
now examine the effects of the vibrational degrees of freedom. One of
us (VR) suggested, from extended H\"uckel calculations, that the
internal vibrational degrees of freedom of the $P\!t_2 (dta)_4$ dimers
---~and in particular the $(dta)$ torsion modes~---must be of
importance in the electronic structure of the
system~\cite{mmx:vR}. These vibrational modes are essentially local
and independent.  Thus, they can be associated with
optical phonons and modeled using Holstein coupling to the electronic
Hamiltonian. The system can thus be reasonably represented by a
two-bands, $3/4$-filled, extended Hubbard-Holstein model. 

The $dta$ torsion modes expected to be important for the
$P\!t_2I(dta)_4$ compound are soft modes with low vibrational
frequencies. The phase diagram of the $3/4$-filled, extended-Hubbard
Holstein model have been extensively studied~\cite{eph}. For low
frequency vibrational modes, and weak to moderate electron-phonon
coupling constant, it exhibits first a Luttinger liquid phase with
renormalized parameters, then, for larger coupling, a metallic phase
characterized by dominant $4k_F$-Charge Density Wave (CDW)
fluctuations and finally, for intermediate coupling, an insulating
$4k_F$-CDW phase, with an exponentially opening gap. The transition
between the different phases is controlled by the ratio between the
hopping integral renormalized by the Franck-Condon factors, and the
electronic repulsion renormalized by the phonons. 

Starting from the $4k_F$-CDW phase, not too far from the phase
transition, and increasing the temperature, one excites the system in
vibrational modes of higher quantum number and thus the
Franck-Condon factors increase. Consequently, the effective hopping integral
increases and the system enters the metallic phase, in agreement with the
$P\!t_2I(dta)_4$ transport measurements.

In this model, the SC phase is characterized as a $4k_F$-CDW, that is
a phase presenting a charge alternation one unit cell out of two. The
$P\!t_2$ dimers can thus be expected to see a charge alternation
$(P\!t_2)^{2.5-\eta+\mu}$, $(P\!t_2)^{2.5-\eta-\mu}$.  Such a dimer
charge alternation is in agreement with the doubling of the $C=S$
vibrational signal seen in IR experiments at $T_{MS}$.

In conclusion, accurate local {\em ab initio} calculations on the
$P\!t_2(dta)_4$ dimer have called in question the possibility of
intra-dimer charge localization in the $MMX$ compounds.  Indeed, in the
$P\!t_2I(dta)_4$ compound, the two $P\!t$ atoms of a dimer are so
strongly covalently bonded ---~due to short $P\!t-P\!t$ distances~---
that intra-dimer charge localization can be considered as
impossible. The only electronic degree of freedom of the dimers,
pertinent for the low-energy physics, is the anti-bonding
${\sigma_d}^*$ orbital which is delocalized on the two $P\!t$
atoms. In consequence the electronic structure of the different phases
of the $P\!t_2I(dta)_4$ compound cannot be understood in terms of
different patterns of charge localization on the platinum atoms ((A)
to (D)).

After a careful analysis of the different low-energy degrees of
freedom, we have proposed a new model, that is able to account for
both the {\em ab initio} results and the experimental ones. This model
involves, in addition to the ${\sigma_d}^*$ dimer orbital, i) the
charge transfer excitation from the $5p_z$ orbital of the iodine and
ii) the vibrational degrees of freedom of the $dta$ intra-dimer
ligands. Such a $3/4$-filled, two-bands, extended-Hubbard Holstein
model can account for the  phase transition, as
well as the XPS and IR experimental results. 

Looking now at similar systems, the present conclusions should hold
for analog $P\!t (RCS_2)_4 I$ ($R=$ethyl, propyl, butyl) compounds,
since, both the intra-dimer distances are very similar (ethyl~:
$2.684$\AA, propyl~: $2.675$\AA~ to $2.689$\AA, butyl~: $2.68$\AA),
and the external part of the intra-dimer ligand does not seem to play
an important role on the electronic parameters~\cite{taku}.

For the $A_4 P\!t_2 (pop)_4 X$ family, the $P\!t-P\!t$ distances are
somewhat larger ($\sim 2.8$\AA) resulting in a smaller
bonding-antibonding intra-dimer splitting. As a consequence, the
resistance of the intra-dimer delocalization to localizing
perturbations should be weaker. However, experimental data do not
exhibit such a localization, except maybe under large pressures where
inter-dimer effects start to compete~\cite{mmx:pop}.  The pertinence of
the charge-transfer degree of freedom from the iodine to the dimer is
also questionable in these systems. Indeed, it is strongly dependent
on the $P\!t-P\!t$ and $P\!t-I$ distances.  A sign that supports the
hypothesis of the non-relevance of this degree of freedom for the
$pop$ systems is the $1:1$, $P\!t^{2+}/P\!t^{3+}$ ratio observed in
the XPS spectra of the $A_4 (P\!t)_2 (pop)_4 X$ compounds.

\begin{acknowledgements} \sf
We would like to thank D. Maynau for providing us with the
CASDI codes and N. Guihery for helpful discussions. 
\end{acknowledgements}

 \end{document}